# Conformational manipulation of DNA in nanochannels using hydrodynamics

Qihao He[1,2], Hubert Ranchon[1,2], Pascal Carrivain[3,4], Yannick Viero[1,2], Joris Lacroix[1,2], Charline Blatché[1,2], Emmanuelle Daran[1,2], Jean-Marc Victor[3,4], Aurélien Bancaud[1,2]

[1] CNRS, LAAS, 7 avenue du colonel Roche, F-31400 Toulouse, France

[2] Univ de Toulouse, LAAS, F-31400 Toulouse, France

[3] LPTMC UMR 7600, CNRS, Universite Pierre et Marie Curie-Paris 6, 4 place Jussieu, 75252 Paris Cedex 05, France

[4] CNRS GDR 3536, Universite Pierre et Marie Curie-Paris 6, 4 place Jussieu, 75252 Paris Cedex 05, France

Correspondence: abancaud@laas.fr

## Abstract

The control over DNA elongation in nanofluidic devices holds great potential for large-scale genomic analysis. So far the manipulation of DNA in nanochannels has been mostly carried out with electrophoresis and seldom with hydrodynamics, although the physics of soft matter in nanoscale flows has raised considerable interest over the last decade. In this report the migration of DNA is studied in nanochannels of lateral dimension spanning 100 to 500 nm using both actuation principles. We show that the relaxation kinetics are 3-fold slowed down and the extension increases up to 3-folds using hydrodynamics. We propose a model to account for the onset in elongation with the flow, which assumes that DNA response is determined by the shear-driven lift forces mediated by the proximity of the channels' walls. Overall we suggest that hydrodynamic actuation allows for an improved manipulation of DNA in nanochannels.

The developments in DNA sequencing technologies have been complemented by innovative methods aiming to obtain genomic maps with a low resolution larger than $10^3$ bp, which can be sufficient to discriminate genomic imprints of individuals[1]. These methods, which have been vaunted for their high-throughput capabilities, generally involve whole chromosome manipulation and structural analysis by optical microscopy at the single molecule level. For instance DNA combing, which consists in spreading DNA molecules on hydrophobic surfaces by applying a receding meniscus,[2, 3] was successfully applied to map the position of genomic sequences along linearized chromosomes, and it enabled to detect microdeletions in genes involved in tuberous sclerosis[4]. The technology of optical mapping allowed to produce whole genome restriction maps in bacteria[5] and more recently in human[6] and in goat[7] based on the sizing of chromosome fragments frozen in a linearly elongated state using an agarose matrix and shear flows.[8]

The throughput of these technologies was further improved with the advent of micro- and nano-fluidics, which enable to generate tailored fluid flows that control the conformation of DNA. For instance elongational flows generated by funnels were shown to induce the transient spreading of DNA, allowing for the genomic mapping of target sequences in bacterial artificial chromosomes of 200 kbp at a throughput of 150 Mbp/s.[9, 10] The conformation of chromosomes was controlled more precisely in nanofluidic channels: wall steric interactions trigger the steady elongation of DNA, and the degree of elongation is determined by the design of nanochannels.[11, 12] This technology was shown to be compatible with DNA restriction mapping,[13] and positioning of target sequences in genomic DNA.[14, 15]

The controlled geometry of nanofluidic devices appeared to be an excellent technology to investigate the physics of DNA in confined environments, which had been thoroughly studied theoretically in the 80's.[16, 17] For instance the degree of DNA elongation was enhanced by decreasing the lateral dimensions of nanochannels,[11] by decreasing the salt concentration,[18, 19] or by adding neutral polymers.[20, 21] In these experiments DNA manipulation was mostly performed using electrophoresis as actuation principle, because this method scales favorably to convey DNA in nanochannels in comparison to hydrodynamics, which requires the use of high pressure sources.[22] Nevertheless the fluid dynamics of polymer solutions in confined geometries raises conceptual challenges, which are essentially associated to the hydrodynamic interactions of polymers with channels' walls.[23] Furthermore the hydrodynamic migration of DNA in narrow capillaries opens new perspectives in analytical sciences, for the principle for matrix-free separation of DNA has recently been demonstrated.[24]

In this report, we investigate the statics and the dynamics of DNA molecules of 50 kbp and ~1 Mbp inside nanochannels of lateral dimension spanning 100 to 500 nm in the presence of hydrodynamics or electrophoresis (Fig. 1A-D). Our study is divided in four sections, which consecutively describe the entry of DNA in nanochannels, its relaxation, and its steady conformation. We first show that DNA molecules nearly systematically fold in a linear conformation with hydrodynamics, whereas they frequently present loops at their leading edge with hydrodynamics. We then measure relaxation kinetics at the single molecule level, and uncover 3-fold slowed down dynamics with hydrodynamic actuation. We subsequently focus on DNA steady extension, and show that the shear constraints imposed by Poiseuille flows enable to control the level of spreading, which can be enhanced up to 3-fold in comparison to electrophoresis. We derive scaling predictions to account for this data, assuming that shear-driven lift forces mediated by the proximity of the channels' walls are responsible for DNA elongation. Overall our study shows that hydrodynamic actuation allows for an improved manipulation of DNA, and paves the way to the rational use of hydrodynamic actuation as an efficient solution for single molecule nanofluidic experiments.

# DNA is preferentially linearly elongated with hydrodynamics

Using our nanofluidic set-up to perform single molecule experiments described in ref. [25] and our custom-made imaging system (see methods section), the entry dynamics of single λ-DNA molecules has been studied in 200 nm square nanochannels. A set point triggering the uptake of DNA has been detected at 20+/-5 bar/cm and 700+/-100 V/cm, and DNA migration velocity 4+/-1 and 20+/-4 µm/s for hydrodynamic and electrophoretic actuation, respectively (dashed line in Fig. 2A). This threshold is consistent with the idea of an entropic barrier that has to be crossed to enter nanochannel,[26] and the solvent flux to trigger the passage of DNA is $k_B T/\eta$ ~$8.10^{-19}$ m$^3$/s given that the viscosity is $\eta$~5 cP, as inferred from Couette rheometry (not shown; $k_B$ is the Boltzmann constant and $T$ the temperature of ~300 K). For a square nanochannel of 200 nm, the mean fluid velocity is thus expected to be 20 µm/s, in excellent agreement with electrophoretic measurements. Above this threshold the velocity of λ-DNA increases approximately linearly with the pressure drop or the electric field (Fig. 2A), enabling us to evaluate the mobility of 3.2 $10^{-4}$ cm$^2$/bar.s and 0.2 $10^{-4}$ cm$^2$/V.s for hydrodynamics and electrophoresis, respectively. Note that the threshold field strength to force the passage of molecules is consistent with earlier studies,[27] but the electrophoretic mobility is slow, and we recently showed that this response was specific to PDMS nanochannels.[25] DNA hydrodynamic mobility compares well with the mobility of a particle travelling at the mean flow velocity in square nanochannels of 200 nm, which is 2.8 $10^{-4}$ cm$^2$/bar.s using the characteristics of Poiseuille flows defined in e.g. ref.[28]. Thus the principles driving DNA uptake and migration can be recapitulated with simple physics models.

The entropic barrier at the interface between micro- and nano-structures causes the trapping of molecules at the entry of nanochannels,[29] and favors the formation of hairpins at the DNA leading edge.[30] Indeed the forces acting on DNA are much greater in nano- than in micro-channels due to the conservation of the electric or hydrodynamic fluxes, so whenever one DNA segment faces the entrance of nanochannels, it drags the molecule, which frequently adopts a looped conformation. We thus set out to evaluate the proportion of hairpins by monitoring the fluorescence intensity along the molecule, given that the leading edge is two-fold brighter for looped DNA (Fig. 2B). The proportion of hairpins is ~85% (n=88) using electrophoresis, and it is nearly constant with the electric field (red dataset in Fig. 2C). This conformation is energetically unfavorable due to the bending and the repulsion of the DNA strands in the loop, but the relaxation towards a linearly elongated state is slow, occurring in tens of seconds (not shown), as was also observed in earlier studies.[14, 30, 31]

When hydrodynamic pressure is applied to force DNA uptake, hairpins are also detected, but they rapidly unfold, leading the formation of DNA molecules linearly elongated conformation (Fig. 2B, upper panel). For slow migration velocities of ~50 µm/s, the vast majority of molecules is linear (hairpin proportion of 5%, blue dataset in Fig. 2C), and the proportion of hairpins increases to 60% at 200 µm/s, a value yet much lower than with electrophoresis. Thus the use of hydrodynamics allows for a better control on the conformation of DNA at the entry of nanochannels. This statement is strongly reinforced by the demonstration that chromosomes fragments of hundreds to thousands of kbp in length can be manipulated in nanochannels with a low proportion of hairpins of 46% (n=52) at a migration velocity of 200 µm/s using hydrodynamics (Fig. 2D).

In conclusion we demonstrate that hydrodynamic actuation allows for a better control over DNA conformation during the uptake in nanochannels, and it can be employed to complement the strategies described in the literature to avoid the formation of hairpins. For instance it has been suggested to wait for the complete relaxation of the molecules in the linear state.[14] This relaxation however occurs in ~50 s for DNA of 150 kbp, and the time scale is expected to increase with the size of DNA, thus leading to excessively long time periods for chromosomal DNA. Nanoposts arrays have also been fabricated ahead of nanochannels in order to disentangle molecules before their uptake,[29] and hence favor the linear state. Interestingly we recently showed that DNA / nanoposts hooking events were different using hydrodynamics *vs*. electrophoretics actuation due to the presence of shear in Poiseuille flows.[32] DNA is indeed more elongated and its conformational space is more restricted, hence better controlled, with hydrodynamics, suggesting that nanofluidic experiments combining hydrodynamics and nanoposts arrays should be conducted in the future to achieve optimal configurational manipulations.

## DNA relaxation is slowed down with hydrodynamics

The behavior of single λ-DNAs has then been studied in the course of their migration, only focusing on linearly elongated molecules (stacked time series represented in Fig. 3A). The degree of elongation is maximal right after the uptake due to the transient intramolecular tension between the leading edge, which is dragged in the nanochannel by the force field, and the lagging end that is located ahead of the entropic barrier, and resists the progression in the channel. The initial longitudinal spreading is enhanced from 11.9 to 19.9 µm with hydrodynamics in comparison to electrophoresis at 100 µm/s (Fig. 3B). This data in turn shows that the end-to-end distance transiently reaches ~90% of the contour length with hydrodynamics given that YOYO-1 stained λ-DNA measures $L\sim22$ µm.[33] A relaxation is then detected (Fig. 3C), and its kinetics is characterized by measuring the end-to-end length of the molecule. In square nanochannels of 200 nm and using electrophoretic actuation, the relaxation time decreases from 0.4 to 0.1 s for migration velocities spanning 30 to 100 µm/s, respectively (Fig. 3C). Notably the retraction kinetics is faster than the characteristic time of length fluctuations at steady state, which is ~1 s in 200 nm square channels.[11] The relaxation time is proportional to the ratio of the friction to the stiffness,[11] which scale as $6\pi\eta l$ (ref. [34]) and $(1 - l/L)^{-3}$, according to the worm like chain response,[35] respectively, with $l$ the molecule length and L its contour length. We therefore expect the retraction kinetics to be accelerated as the elongation of the molecule increases, in agreement with our experiments.

DNA relaxation has then been examined with hydrodynamics, showing that the kinetics is three-fold slowed down in comparison to electrophoresis (Fig. 3C). This difference in relaxation dynamics is likely associated to the shear constraints induced by Poiseuille flows, which are characterized by the shear rate $\dot\gamma$ of $2v/h$ ~$10^3$ s$^{-1}$ for a flow velocity $v$ of 100 µm/s in square nanochannels of $h$~200 nm. It has indeed been observed in bulk shear flows that the fluctuations of DNA end-to-end distance are slowed down as the shear rate increases.[33] The same trend has been reported for end-tethered DNA chains in shear flows,[36] and this response was shown to arise from cyclic dynamics that occur due to the coupling between the chain velocity in the flow direction to fluctuations in the shear-gradient direction. Interestingly the observation of slowed down relaxations may account for the larger maximal elongation with hydrodynamics through a shear-enhanced "deformability" of DNA. Indeed the threshold flow to trigger the deformation of an end-tethered polymer is determined by the ratio of its gyration radius to its relaxation time,[37] so slowly relaxing polymers can be extended with low flow rates, or equivalently reach large extensions for a given flow rate. Similarly one may speculate that the linearization of molecules with hairpins is accelerated by the shearing that accelerates the relaxation the stable conformation (Fig. 2B).

Overall we observe that the kinetics of DNA relaxation is slowed down with hydrodynamics, and we propose that the conformational properties of DNA, which are characterized by a higher extension and a lower proportion of hairpins with hydrodynamics, are determined by this change in dynamics.

## The elongation of DNA is controlled by the flow

The elongation of λ-DNA at the end of the relaxation has finally been measured (Fig. 4A-B). The steady extension using electrophoresis is 5.3 µm, hence ~24% of the total length, in square nanochannels of 200 nm, in agreement with earlier studies carried out in the absence of force fields.[11, 25] Thus the electrophoretic actuation has marginal effects on DNA conformation, as expected from the fact that DNA behaves as a free-draining polymer during electrophoresis.[38] Conversely DNA end-to-end distance is more than doubled to ~55% for a migration velocity of 100 µm/s using hydrodynamics (Fig. 4B). This result is reminiscent of the onset in elongation of ~30% in bulk shear flows characterized by $\dot{\gamma}$~5 s$^{-1}$,[33] but the extension exhibited large fluctuations due to an end-over-end tumbling of DNA, which limited the use of this principle for biomolecule analysis (see discussion in e.g. ref. [39]). Tumbling is impeded in nanochannels due to the level of confinement, accounting for the steady elongation in Fig. 4B. In addition the degree of elongation can be monitored from 24% to 65% by tuning the migration velocity in the range 0-200 µm/s (Fig. 4C), allowing us to reproduce the performances of confined nanochannels of ~80 nm in cross-section without changing fabrication technology.[11] We thus argue that the control of the pressure source provides an efficient macroscopic control parameter for the manipulation of single DNA molecules in nanofluidic channels.

**Shear-driven lift forces are responsible for DNA elongation**

We set out to investigate the physics driving the flow-driven elongation of DNA. The end-to-end length of single λ-DNA molecules was monitored in nanochannels of different geometries as specified in Fig. 1D, and for different migration velocities (Fig. 5A), showing a variety of responses that we wished to reconcile with a model. We first considered that DNA behaved as a free-draining polymer, *i.e.* neglecting hydrodynamic interactions between monomers, as was suggested to model critical flow rates for dragging linear polymer in nanochannels.[40, 41] We performed Langevin dynamics simulations of freely-jointed chains confined in nanochannels of 200 nm, and measured the elongation of the molecule in plug or Poiseuille flows. Given the computational cost limitations of simulations, we focused on DNA chains of 1 and 4 µm in contour length, which adopt globular conformations of ~300 and ~600 nm in solution according to our simulations (not shown). The end-to end distance remained roughly constant to ~20% of the contour length for every flow spanning 0-200 µm/s (Supplementary Fig. S2), thus strongly suggesting that this model is inadequate to reproduce the spreading forced by the flow. We then considered that the response of the molecule was determined by its elasticity, as well as monomer/monomer and monomer/wall hydrodynamic interactions. The effect of hydrodynamic interactions has been investigated in molecular dynamics simulations, showing that DNA migrates toward the centerline of the channel where it is stretched.[42] This transverse migration is associated to the development of a depletion layer near the walls, which has been studied numerically, analytically,[43, 44] and experimentally in microchannels.[45] The thickness of the depletion layer is unfortunately not measurable in nanochannels, so we developed a model focused on the response of polymer to shear stress.

We consider that the de Gennes-Pincus blob picture[16] can be applied to study the response of polymers in hydrodynamic flows. According to this model the chain consists of a series of self-avoiding blobs of diameter $D$, each of them behaving as a chain in good solvent, and its degree of elongation is:

$$\frac{l}{L} \sim \left\{\frac{wP}{D^2}\right\}^{1/3} \tag{1}$$

with $w$ and $P$ the polymer diameter and Kuhn length, respectively. The long-distance hydrodynamic coupling between monomers inside each blob should lead to coils behaving as massive objects impermeable to the flow, as observed in e.g. analytical centrifugation (Fig. 5B).[16] We may thus model each blob as a globule, and consider that hydrodynamic interactions with the walls can be treated as in the physics of globules in shear flows. Deformable objects, such as globules, conveyed by flows are indeed subjected to lift forces at the proximity of surfaces, which are expressed at the scaling level as $\eta\dot{\gamma}D^3/y$ with $y$ the distance of the object to the surface.[46] The cumulative effect of each channel wall leads to a net centripetal lift force that tends to elongate the molecule at the expense of an entropic cost. The onset in elongation associated to the flow $\Delta l$ can then be derived (see details of the calculation in supplementary materials):

$$\frac{\Delta l}{L} h^{-8/15} \sim (wP)^{1/3} \left(\frac{\eta}{8k_BT}\right)^{2/5} \dot{\gamma}^{2/5} \tag{2}$$

The numerical prefactor in the right term is 43 m$^{-8/15}$s$^{-2/5}$ taking 2 and 100 nm for the DNA radius and Kuhn length, respectively, showing good correspondence of our model with the data (solid curve in Fig. 5C). Altogether we propose that DNA spreading is forced by the shear, and we argue that our model is readily suited to define operating conditions according to specifications on the degree of spreading required for single molecule genomic analysis.

## Conclusion

We have investigated the migration of DNA in nanochannels using hydrodynamics, and report that relaxation dynamics are slowed down and the steady elongation is enhanced in comparison to electrophoresis. We propose that these changes are associated to the shear constraints imposed by Poiseuille flows and to the existence of hydrodynamic interactions of the polymer with channels' walls. We also derive an analytical model that reproduces the variations of the elongation of the molecule with the flow velocity. Altogether we believe that this report can serve as a guideline to design single molecule manipulation experiments in nanochannels. Future work is also needed to study the response of DNA in nanochannels with lateral dimension narrower than 100 nm, because the physics of the molecule is described by the Odijk regime,[11, 17] which readily departs from the de Gennes response that is relevant to our experimental settings. In another direction molecular dynamics simulations of polymers flowing in micro- and nano-channels have been mostly focused on the thickness of the depletion layer near the walls. The depletion layer is expected to force the spreading of DNA, but the degree of elongation is poorly reproduced by the scaling analysis derived from this model (Supplementary Fig. S3). Further clarifications to reconcile the mechanics of DNA and its depletion from the walls can be obtained from simulations to improve our description of the fluid dynamics of DNA in confined geometries.

## Experimental

### Fabrication

PDMS chips were obtained by the sequential baking of one 40 µm layer hard PDMS covered by a layer of conventional PDMS of ~1 cm for 45 min and 3 hours at 75°C, respectively (see details in reference [25] on the mold fabrication process). Note that PDMS is deformable, but the deformation $\delta$ for square nanochannels of lateral dimension $h$ scales as $\delta/h \sim \Delta P/E$[47] with $E$ the Young modulus of hard PDMS of ~5 MPa[48], implying that $\delta$ is small in our experiments of ~2 nm for an actuation of 0.5 bar. Silicon nanochannels were obtained by electron beam lithography, followed by reactive ion etching using an Alcatel system (etching rate=22 nm/s). Microfluidic channels were produced by conventional photolithography, and access holes, which served as fluidic inlet and outlet, were drilled through silicon by sand blasting. Note that the comparison of electrophoretic *vs.* hydrodynamic actuation required the growth of an insulating layer of silicon dioxide of 100 nm. PMDS channels were eventually sealed after plasma activation of the chip and one glass coverslip, and the sealing of silicon nanochannels was obtained using a thin layer of PDMS spin-coated on glass coverslips (see reference [25] for details of the protocol), that was also activated by oxygen plasma. The bonding strength was enhanced by curing the resulting chips at 100°C during 20 minutes, enabling us to apply pressures up to 2 bars.

### DNA preparation, manipulation, and imaging

DNA was fluorescently labeled with YOYO-1 (Molecular Probes) at a staining ratio of 1 fluorophore per 10 bp after careful titration of DNA and YOYO-1 by absorbance spectroscopy at 260 nm and 488 nm, respectively. The buffer was 1X TBE (89 mM Tris-borate and 2 mM EDTA, pH 8.3), and it was supplemented with 5% Dithiothreitol to reduce photo-induced damages, and with 2% Poly-vinylpyrrolydone (PVP, 40 kDa) to suppress non specific interactions and reduce electro-osmotic flows. Single molecule experiments were conducted with λ-DNA molecules (48.5 kbp), and with genomic DNA obtained from NRK cells. Genomic DNA purification consisted in harvesting $10^6$ cells and resuspending the cells in 2 mL of a solution containing 1% low-melt agarose. This mix was then dispensed in 10 scaffolds of 10x10x2 mm$^3$ until agarose reticulated, and these agarose blocks used as matrices to preserve genomic DNA from mechanical constraints during purification, which was performed using 1% SDS and 250 µg/mL proteinase K. Agarose blocks were eventually stored in TE (Tris-HCl 10 mM and EDTA 1 mM) at 4°C. They were melted at 42°C with 1unit/mL β-agarase before experiments.

Imaging was performed with a Zeiss epifluorescence microscope equipped with the 38HE filter set (Zeiss), and with a Lumencor Light Engine emitting at 475 nm with a 28 nm bandwidth and a power of 20 mW. An ANDOR iXon-885 camera was used to observe single DNAs using a binning of 4*4, and a pixel size of 0.33 µm. Exposure times varied from 20 to 40 ms depending on the requirements of the experiment (see Supplementary Video 1 to assess the quality of videos). DNA manipulation was performed using electrophoresis or pressure actuation that was monitored by a Fluigent pressure manager in the range 10 to 1000 mBar. The pressure was applied uniformly in the two inlet and outlet channels (Fig. 1A). Note that this technology provides a stable and pulseless flow with a rapid response in comparison to syringe pumps, which deliver hysteretic flows with long equilibration times for nanoscale volumes.

Videos sequences were eventually analyzed using custom macros implemented in ImageJ. Molecules were automatically segmented by Otsu thresholding in order to retrieve its center of mass and length at every time step.

## Acknowledgements:

We are grateful to F. Brochard for enlightening discussions on the manuscript. This work and the fellowship of Q. He were supported by the ANR program JC08_341867. This work was partly supported by the LAAS-CNRS technology platform, a member of the French RENATECH network. Y. Viero and J. Lacroix thank the Defense Procurement Agency (DGA) and the RITC foundation for PhD fellowship funding, respectively.

**Figure 1: Nanofluidic devices for hydrodynamic and electrophoretic DNA manipulation.**
**(A)** The scheme represents the fluidic chip, which consists of an array of nanochannels connected to two microfluidic channels and four macroscopic inlets/outlets. The photograph in the right shows the functional device, and we monitor the electric field and/or the pressure in each reservoir. The flow profile is parabolic with hydrodynamics, and nearly flat with electrophoresis given that the Debye layer thickness represents ~1 nm in 1X TBE. **(B)** Scanning electron micrographs (SEM) of a silicon mold composed of square nanochannels of 200 nm, and the replicate of this structure in hPDMS (see methods and ref [25], scale bars = 2 µm). **(C)** The electron micrographs represent the different silicon nanochannels, their widths spanning 100 nm to 500 nm (scale bars = 200 nm). **(D)** The table specifies the geometry of the nanochannels used in this study.

**Figure 2: Uptake of DNA in nanochannels using hydrodynamics and electrophoresis. (A)** The velocity of λ-DNA molecules is plotted as a function of the pressure drop or electric field (left and right panel, respectively), showing the existence of a threshold (dashed lines) above which the migration velocity increases roughly linearly with the actuation. **(B)** The two time series show the uptake of λ-DNA in PDMS square nanochannels of 200 nm using an hydrodynamic or electrophoretic actuation (upper and lower panels, respectively). **(C)** The plot represents the number of molecules with an extended conformation, as shown in the inset, after a migration of 20 µm in the nanochannel as a function of the migration velocity, when hydrodynamics or electrophoresis is used for actuation (blue and red datasets, respectively). **(D)** Genomic DNA is also efficiently manipulated with hydrodynamics, as shown by the time series of the migration of one chromosome fragment entering nanochannels with hydrodynamics at a migration velocity of 200 µm/s. The residence time of the molecule is ~2.7 s, so we estimate that its length is ~550 µm, or equivalently ~2 Mb, given that λ-DNA elongation is ~15 µm (see Fig. 4 below).

**Figure 3: λ-DNA relaxation dynamics. (A)** The stacked time series of micrographs shows the migration of one λ-DNA in a square nanochannel of 200 nm at a velocity of 30 µm/s using electrophoresis. The upper image is recorded as the molecule has just completely entered the nanochannel, and the time interval between consecutive images is 45 ms. **(B)** The histogram represents λ-DNA maximal extension, *i.e.* right after their uptake in nanochannels, using electrophoretic or hydrodynamic actuation. The migration velocity is set to 100 µm/s. **(C)** The upper panel shows the time series of (A) after the registration of the lagging end of the molecule in order to visualize the relaxation dynamics. The plot compares the relaxation time, as inferred from single exponential fitting of end-to-end length measurements (see Supplementary Fig. S1), as a function of the migration velocity in the case of electrophoretic or hydrodynamic actuation. Each data point is an average measurement over at least three molecules.

**Figure 4: Flow-driven DNA elongation in nanofluidic channels. (A)** The upper and lower panels are stacked time series of single λ-DNA molecules in the course of their migration at 100 µm/s using electrophoresis or hydrodynamics. Note that the lagging edges of the molecules are registered, as described in Fig. 3C. **(B)** The histogram shows the steady extension using electrophoretic and hydrodynamic actuation (red and blue datasets, respectively) at a migration velocity set to 100 µm/s. **(C)** The plot represents the steady

extension of single λ-DNA molecules as a function of the migration velocity using hydrodynamic actuation. The extension was constant with the electric field (dashed line).

**Figure 5: Scaling analysis of flow-driven elongation. (A)** The steady extension of single λ-DNA molecules was recorded as a function of their migration velocity for different nanochannels geometries, as indicated in the inset. Each data point is an average measurement over at least 5 molecules. **(B)** Schematic representation of our model describing DNA elongation: the chain is divided in blobs, in which the flow profile (blue line) is flat due to hydrodynamic interactions between polymer segments. The shear flow near the walls exerts a centripetal lift force on each blob that tends to elongate the molecule. **(C)** The datasets of Fig. 4C and 5A are rescaled using the shear rate on the x-axis, which is defined by $v_m\{1/a + 1/b\}$ with $v_m$ the molecule velocity, and $a$ and $b$ the channel width and height, respectively. The y-axis is the normalized variation in extension corrected by the channel lateral dimension, as specified in the de Gennes model $h = \sqrt{ab}$. The bold line corresponds to the prediction of our model (Eq. (2)).

# Bibliography

1.      Lam, E. T.; Hastie, A.; Lin, C.; Ehrlich, D.; Das, S. K.; Austin, M. D.; Deshpande, P.; Cao, H.; Nagarajan, N.; Xiao, M.; Kwok, P.-Y. *Nat Biotech* **2012,** 30, (8), 771-776.

2.      Bensimon, A.; Simon, A.; Chiffaudel, A.; Croquette, V.; Heslot, F.; Bensimon, D. *Science* **1994,** 265, 2096-2098.

3.      Bensimon, D.; Simon, A. J.; Croquette, V.; Bensimon, A. *Phys Rev Lett* **1995,** 74, 4754-4757.

4.      Michalet, X.; Ekong, R.; Fougerousse, F.; Rousseaux, S.; Schurra, C.; Hornigold, N.; van Slegtenhorst, M.; Wolfe, J.; Povey, S.; Beckman, J. S.; Bensimon, A. *Science* **1997,** 277, 1518-1523.

5.      Lin, J.; Qi, R.; Aston, C.; Jing, J.; Anantharaman, T. S.; Mishra, B.; White, O.; Daly, M. J.; Minton, K. W.; Venter, J. C.; Schwartz, D. C. *Science* **1999,** 285, 1558-1562.

6.      Teague, B.; Waterman, M. S.; Goldstein, S.; Potamousis, K.; Zhou, S.; Reslewic, S.; Sarkar, D.; Valouev, A.; Churas, C.; Kidd, J. M.; Kohn, S.; Runnheim, R.; Lamers, C.; Forrest, D.; Newton, M. A.; Eichler, E. E.; Kent-First, M.; Surti, U.; Livny, M.; Schwartz, D. C. *Proc Nat Acad Sci USA* **2010,** 107, (24), 10848-10853.

7.      Dong, Y.; Xie, M.; Jiang, Y.; Xiao, N.; Du, X.; Zhang, W.; Tosser-Klopp, G.; Wang, J.; Yang, S.; Liang, J.; Chen, W.; Chen, J.; Zeng, P.; Hou, Y.; Bian, C.; Pan, S.; Li, Y.; Liu, X.; Wang, W.; Servin, B.; Sayre, B.; Zhu, B.; Sweeney, D.; Moore, R.; Nie, W.; Shen, Y.; Zhao, R.; Zhang, G.; Li, J.; Faraut, T.; Womack, J.; Zhang, Y.; Kijas, J.; Cockett, N.; Xu, X.; Zhao, S.; Wang, J.; Wang, W. *Nat Biotech* **2013,** 31, (2), 135-141.

8.      Schwartz, D. C.; Li, X. H., L. l.; Ramnarain, S. P.; Huff, E. J.; Wang, Y. K. *Science* **1993,** 262, 110-114.

9.    Chan, E. Y.; Goncalves, N. M.; Haeusler, R. A.; Hatch, A. J.; Larson, J. W.; Maletta, A. M.; Yantz, G. R.; Carstea, E. D.; Fuchs, M.; Wong, G. G.; Gullans, S. R.; Gilmanshin, R. *Genome Res* **2004,** 14, (6), 1137-1146.

10.    Phillips, K. M.; Larson, J. W.; Yantz, G. R.; D'Antoni, C. M.; Gallo, M. V.; Gillis, K. A.; Goncalves, N. M.; Neely, L. A.; Gullans, S. R.; Gilmanshin, R. *Nucl Acid Res* **2005,** 33, (18), 5829-5837.

11.    Reisner, W. W.; Morton, K. J.; Riehn, R.; Wang, Y. M.; Yuan, Z.; Rosen, M.; Sturm, J. C.; Chou, S. Y.; Frey, E.; Austin, R. H. *Physical Review Letters* **2005,** 94, 196101.

12.    Tegenfeldt, J. O.; Prinz, C.; Cao, H.; Chou, S. Y.; Reisner, W. W.; Riehn, R.; Wang, Y. M.; Cox, E. C.; Sturm, J. C.; Silberzan, P.; Austin, R. H. *Proc Natl Acad Sci USA* **2004,** 101, 10979-10983.

13.    Riehn, R.; Lu, M.; Wang, Y.-M.; Lim, S. F.; Cox, E. C.; Austin, R. H. *Proc Natl Acad Sci USA* **2005,** 102, (29), 10012-10016.

14.    Jo, K.; Dhingra, D. M.; Odijk, T.; de Pablo, J. J.; Graham, M. D.; Runnheim, R.; Forrest, D.; Schwartz, D. C. *Proc Natl Acad Sci USA* **2007,** 104, (8), 2673-2678.

15.    Das, S. K.; Austin, M. D.; Akana, M. C.; Deshpande, P.; Cao, H.; Xiao, M. *Nucl Acid Res* **2010,** 38, (18), e177.

16.    de Gennes, P.-G., *Scaling concepts in polymer physics*. Cornell university press: Ithaca, 1979.

17.    Odijk, T. *Macromolecules* **1983,** 16, (8), 1340-1344.

18.    Kim, Y.; Kim, K. S.; Kounovsky, K. L.; Chang, R.; Jung, G. Y.; dePablo, J. J.; Jo, K.; Schwartz, D. C. *Lab on a Chip* **2011,** 11, (10), 1721-1729.

19.    Reisner, W.; Beech, J. P.; Larsen, N. B.; Flyvbjerg, H.; Kristensen, A.; Tegenfeldt, J. O. *Physical Review Letters* **2007,** 99, (5), 058302.

20.     Jones, J. J.; van der Maarel, J. R. C.; Doyle, P. S. *Nano Letters* **2011,** 11, (11), 5047-5053.

21.     Zhang, C.; Shao, P. G.; van Kan, J. A.; van der Maarsel, J. R. C. *Proc Natl Acad Sci USA* **2009,** 106, 16651-16656.

22.     Reisner, W.; Larsen, N. B.; Silahtaroglu, A.; Kristensen, A.; Tommerup, N.; Tegenfeldt, J. O.; Flyvbjerg, H. *Proc Natl Acad Sci USA* **2010,** 107, (30), 13294-13299.

23.     Graham, M. D. *Annu Rev Fluid Mech* **2011,** 43, 273-298.

24.     Wang, X.; Liu, L.; Guo, G.; Wang, W.; Liu, S.; Pu, Q.; Dasgupta, P. K. *TrAC Trends in Analytical Chemistry* **2012,** 35, (0), 122-134.

25.     Viero, Y.; He, Q.; Mazenq, L.; Ranchon, H.; Fourniols, J. Y.; Bancaud, A. *Microfluidics Nanofluidics* **2012,** 12, 465-473.

26.     de Gennes, P.-G., Flexible polymers in nanopores. In *Advances in Polymer Science*, Springer-Verlag, Ed. Berlin, 1999; Vol. 138, pp 91-105.

27.     Menard, L. D.; Ramsey, J. M. *Analytical Chemistry* **2013,** 85, (2), 1146-1153.

28.     Stone, H. A., Introduction to Fluid Dynamics for Microfluidic Flows. In *CMOS Biotechnology*, Lee, H.; Ham, D.; Westervelt, R. M., Eds. Springer: Cambridge, MA, 2007; p 379.

29.     Cao, H.; Tegenfeldt, J. O.; Austin, R. H.; Chou, S. Y. *Applied Physics Letters* **2002,** 81, (16), 3058-3060.

30.     Reccius, C. H.; Stavis, S. M.; Mannion, J. T.; Walker, L. P.; Craighead, H. G. *Biophysical Journal* **2008,** 95, (1), 273-286.

31.     Mannion, J. T.; Reccius, C. H.; Cross, J. D.; Craighead, H. G. *Biophysical Journal* **2006,** 90, (12), 4538-4545.

32.     Viero, Y.; He, Q.; Bancaud, A. *Small* **2011,** 7, 3508–3518.

33.     Smith, D. E.; Babcock, H. P.; Chu, S. *Science* **1999,** 283, 1724-1727.

34.     Brochard, F.; De Gennes, P.-G. *J Chem Phys* **1977,** 67, 52-56.

35.     Perkins, T. T.; Quake, S. R.; Smith, S. B.; Chu, S. *Science* **1994,** 264, 822-826.

36.     Doyle, P. S.; Ladoux, B.; Viovy, J. L. *Physical Review Letters* **2000,** 84, (20), 4769-4772.

37.     Brochard-Wyard, F. *Europhys. Lett.* **1995,** 30, (7), 387-392.

38.     Viovy, J. L. *Rev. Mod. Phys.* **2000,** 72, (3), 813-872.

39.     Larson, J. W.; Yantz, G. R.; Zhong, Q.; Charnas, R.; D'Antoni, C. M.; Gallo, M. V.; Gillis, K. A.; Neely, L. A.; Phillips, K. M.; Wong, G. G.; Gullans, S. R.; Gilmanshin, R. *Lab on a Chip* **2006,** 6, (9), 1187-1199.

40.     Freed, K. F.; Wu, C. *J Chem Phys* **2011,** 135, 144902-144908.

41.     Freed, K. F.; Wu, C. *Macromolecules* **2012,** 44, 9863-9866.

42.     Jendrejack, R. M.; Dimalanta, E. T.; Schwartz, D. C.; Graham, M. D.; de Pablo, J. J. *Phys Rev Lett* **2003,** 91, 038102-1.

43.     Ma, H.; Graham, M. D. *Phys Fluids* **2005,** 17, 083103.

44.     Chen, Y. L.; Graham, M. D.; de Pablo, J. J.; Jo, K.; Schwartz, D. C. *Macromolecules* **2005,** 38, 6680-6687.

45.     Jo, K.; Chen, Y.-L.; de Pablo, J. J.; Schwartz, D. C. *Lab Chip* **2009,** 9, 2348-2355.

46.     Abkarian, M.; Lartigue, C.; Viallat, A. *Phys Rev Lett* **2002,** 88, 068103.

47.     Gervais, T.; El-Ali, J.; Günther, A.; Jensen, K. F. *Lab on a Chip* **2006,** 6, 500-507.

48.     Schmid, H.; Michel, B. *Macromolecules* **2000** 33, 3042–3049.

Fig. 1

He *et al.*

A

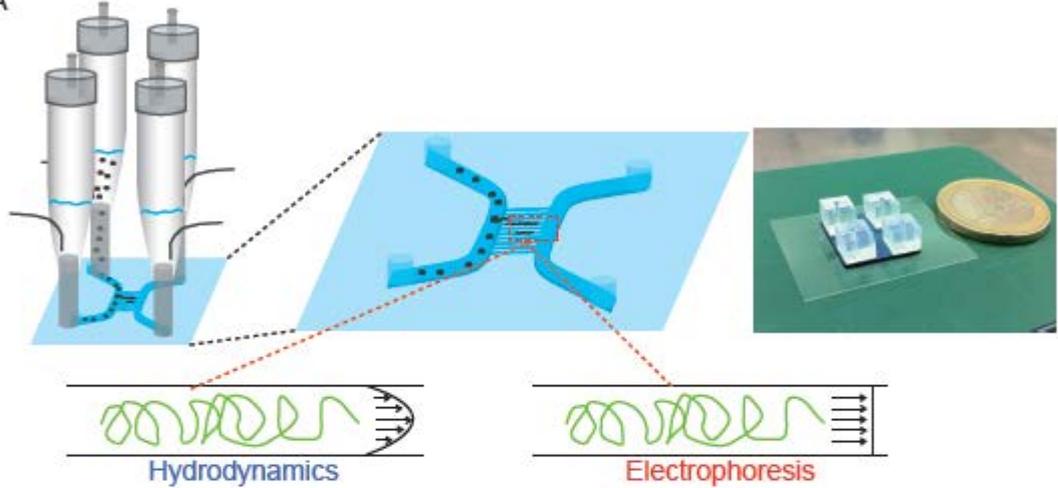

Hydrodynamics          Electrophoresis

B

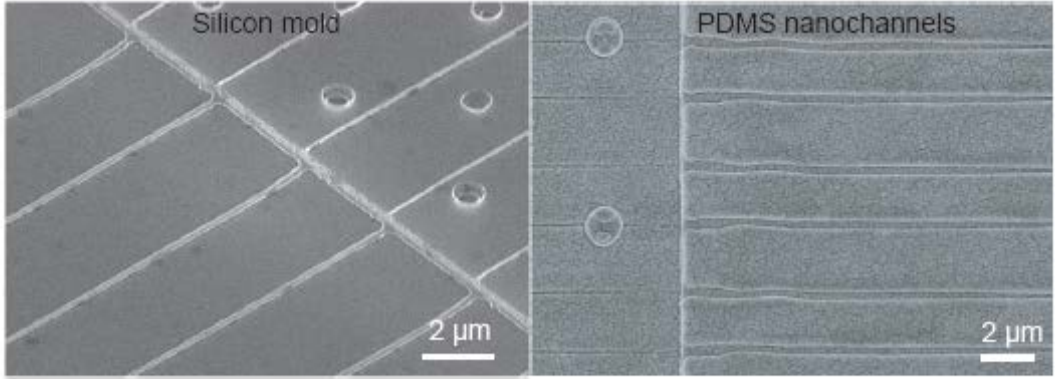

C

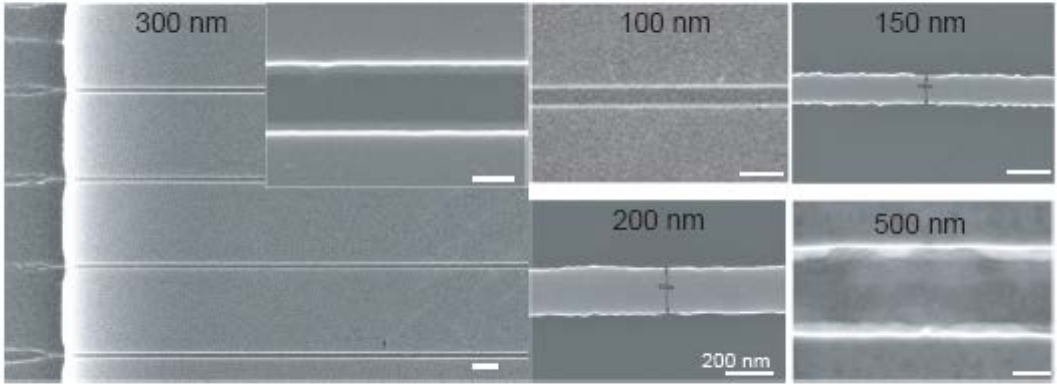

D

| Material | Width (nm) | Height (nm) | Length (µm) |
|----------|-----------|-------------|-------------|
| PDMS | 200 | 200 | 100 |
| Silicon | 100 | 200 | 80 |
| Silicon | 150 | 200 | 100 |
| Silicon | 300 | 250 | 100 |
| Silicon | 500 | 250 | 500 |

He *et al.* (2012)

Fig. 2                                                                                                    He *et al.*

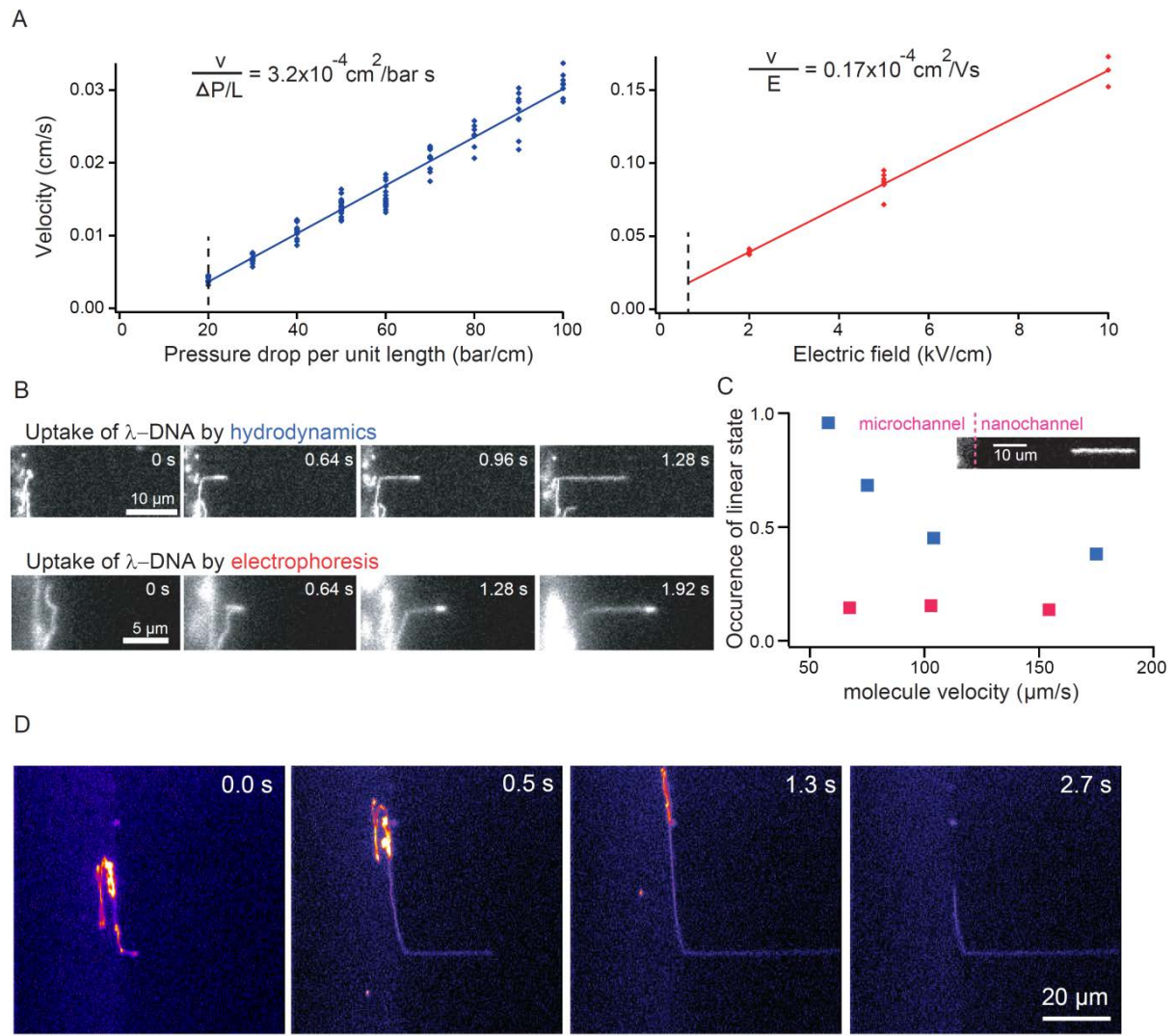

A

$$\frac{v}{\Delta P/L} = 3.2 \times 10^{-4}\ \text{cm}^2/\text{bar s}$$

$$\frac{v}{E} = 0.17 \times 10^{-4}\ \text{cm}^2/\text{Vs}$$

B

Uptake of λ–DNA by hydrodynamics
0 s    0.64 s    0.96 s    1.28 s
10 µm

Uptake of λ–DNA by electrophoresis
0 s    0.64 s    1.28 s    1.92 s
5 µm

C
microchannel   nanochannel
10 um

D
0.0 s    0.5 s    1.3 s    2.7 s
20 µm

Fig. 3                                        He et al.

A

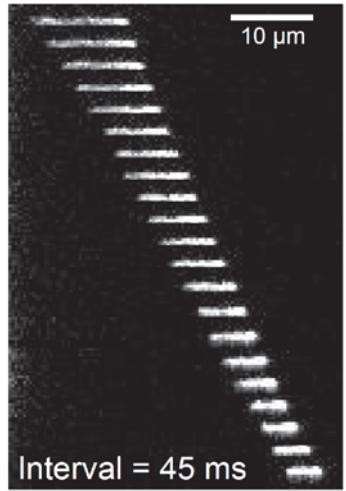

B

Maximal extension at v=100 µm/s

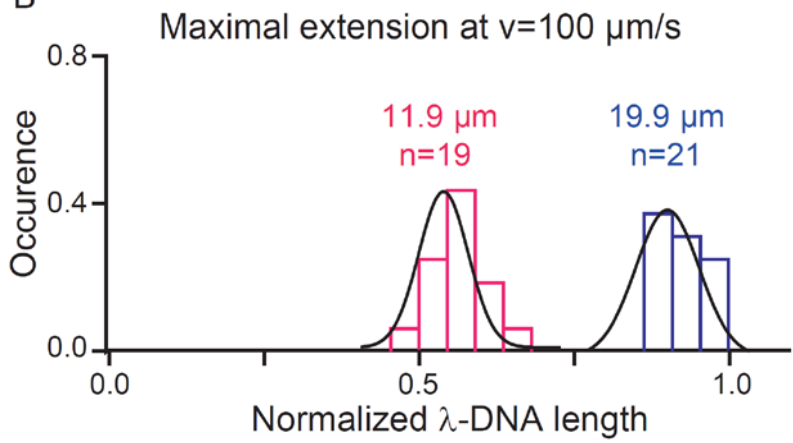

C

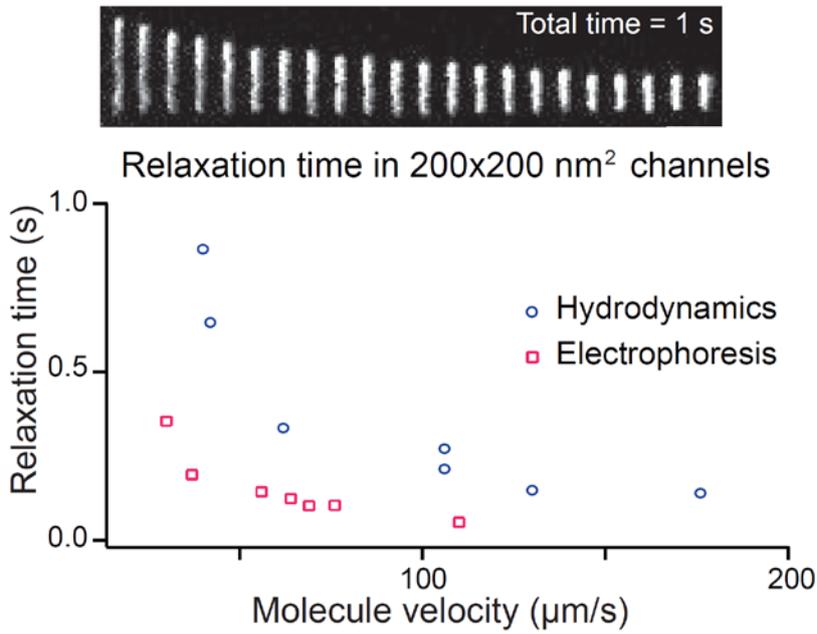

Fig. 4                                          He *et al.*

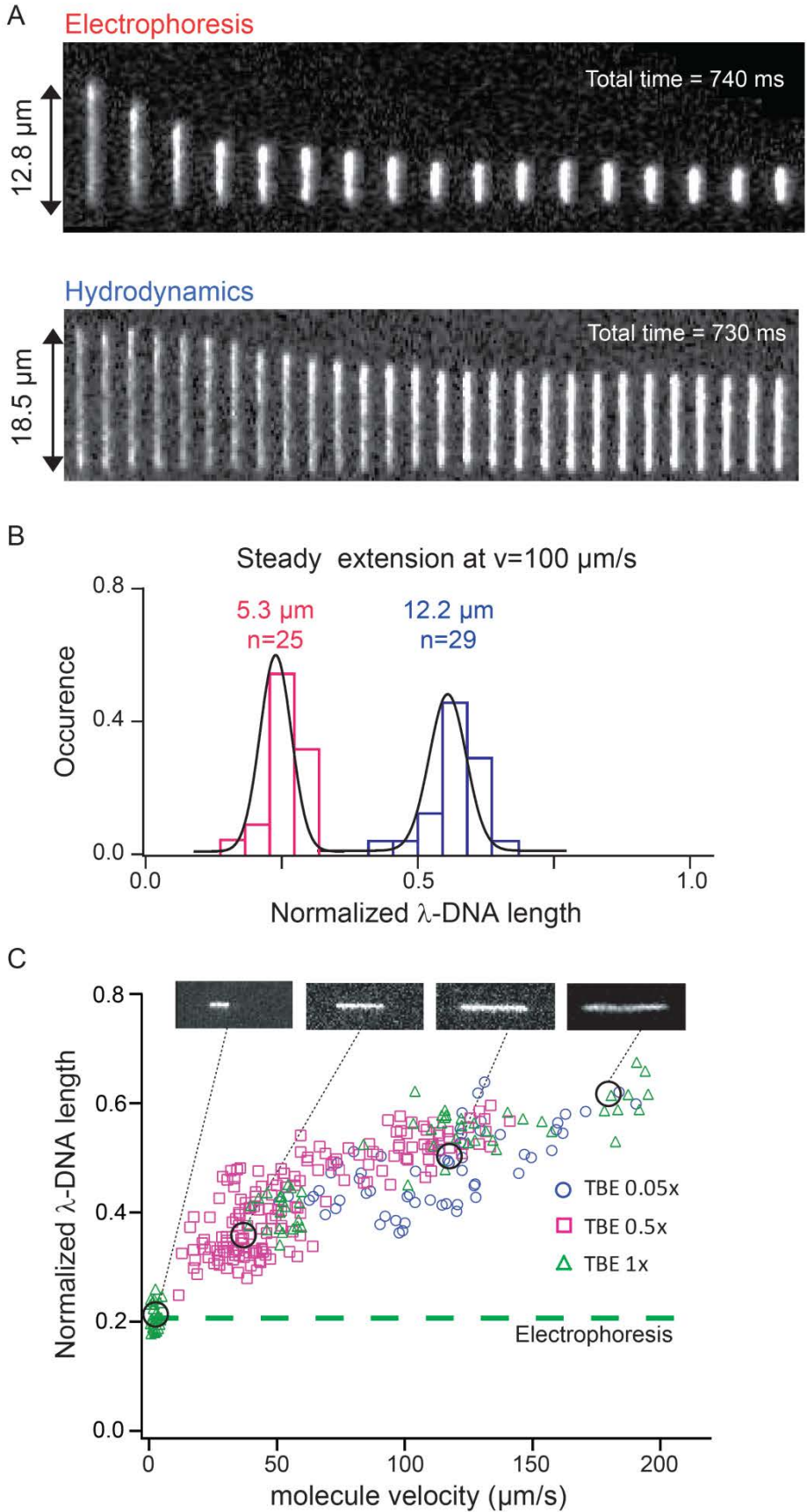

A  Electrophoresis

Total time = 740 ms

12.8 µm

Hydrodynamics

Total time = 730 ms

18.5 µm

B

Steady extension at v=100 µm/s

5.3 µm
n=25

12.2 µm
n=29

Occurence

Normalized λ-DNA length

C

Normalized λ-DNA length

○ TBE 0.05x
□ TBE 0.5x
△ TBE 1x

Electrophoresis

molecule velocity (µm/s)

Fig. 5                                    He *et al.*

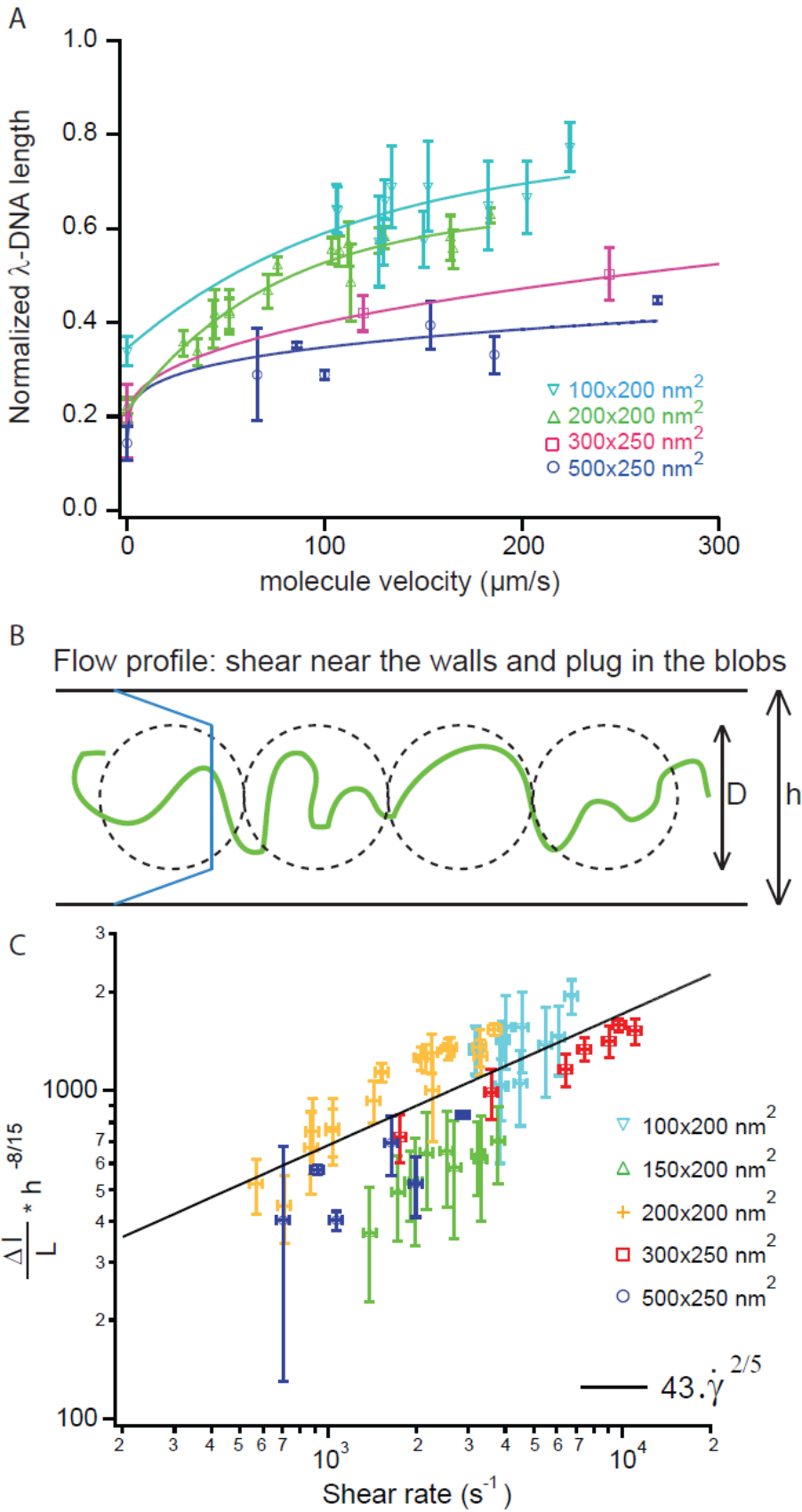

A

B — Flow profile: shear near the walls and plug in the blobs

C